\documentclass[12pt]{article}
\thicklines
\newcommand{\EQ}{\begin{equation}}
\newcommand{\EN}{\end{equation}}
\newcommand{\bea}{\begin{eqnarray}}
\newcommand{\ena}{\end{eqnarray}}
\newcommand{\eea}{\end{eqnarray}}

\def\3i{\int\!\!\!\int\!\!\!\int}
\def\2i{\int\!\!\!\int}

\def\sqr#1#2{{\vcenter{\vbox{\hrule height.#2pt
     \hbox{\vrule width.#2pt height#1pt \kern#1pt
           \vrule width.#2pt}
       \hrule height.#2pt}}}}
\def\square{\mathchoice\sqr55\sqr55\sqr{2.1}3\sqr{1.5}3}
\begin{document}

\begin{flushright}
\begin{minipage}{0.25\textwidth} hep-th/0104234
\end{minipage}
\end{flushright}
\begin{center}
\bigskip\bigskip\bigskip
{\bf\Large{
Simplified Method for Trace Anomaly Calculations 
 \\ \vskip .2cm  
in $d \leq 6 $
}}
\vskip 1cm
\bigskip
Fiorenzo Bastianelli $^a$\footnote{E-mail: bastianelli@bo.infn.it} 
and
N.D. Hari Dass $^b$\footnote{E-mail: dass@imsc.ernet.in} 
\\[.4cm]
{\em $^a$ Dipartimento  di Fisica, Universit\`a di Bologna 
and  INFN, Sezione di Bologna\\ 
via Irnerio 46, I-40126 Bologna, Italy} \\[.4cm]
{\em $^b$ Institute of Mathematical Sciences, C.I.T. Campus, 
Chennai 600 113, India}\\
\end{center}
\baselineskip=18pt
\vskip 2.3cm

\centerline{\large{\bf Abstract}}
\vspace{.4cm}

We discuss a simplified method for computing trace anomalies
in dimensions $d \leq 6 $. It is known that in the quantum mechanical 
approach trace anomalies in $d$ dimensions are given by a 
$({d\over 2}+1)$-loop computation in an auxiliary 1d sigma model with 
arbitrary geometry. We show how one can obtain the same information
using a simpler ${d\over 2}$-loop calculation on an arbitrary geometry
supplemented by a $({d\over 2} +1)$-loop calculation on the simplified 
geometry of a maximally symmetric space.

\newpage


Conformal anomalies, also called trace anomalies have a variety of 
applications and are of central importance
in string and quantum field theories 
\cite{duff}. They arise due to the fact that the regularisation
procedure brings in a scale dependence even though the classical
theory was scale invariant.
In particular, they appear whenever conformal field theories in even dimensions
are considered in a curved 
background\footnote{More general background 
field configurations are also usefully considered sometimes.}.
However their computation gets more involved as the spacetime 
dimension $d$ gets bigger.
Already for $d=6$ the computation is quite complicated and 
only recently the anomalies for free conformal fields 
have been completely identified \cite{Bastianelli:2000hi}
(we recall also that
the trace anomalies for a class of supersymmetric  
$(0,2)$ interacting CFT$_6$ in the large $N$ limit was obtained in 
\cite{HS}
using the supergravity dual as dictated by the AdS/CFT correspondence 
\cite{MW}).
Trace anomalies can be computed efficiently using the quantum 
mechanical approach of
\cite{Bastianelli:1992be,Bastianelli:1993ct},
which requires a $(n+1)$-loop calculation
in an auxiliary 1d nonlinear sigma model
to obtain the trace anomalies in $d=2n$ dimensions.
The complications for $d=6$ trace anomalies are seen in this 
method as the need
for calculating up to 4 loops in the 1d sigma model. 
The latter is laborious even in the newly developed dimensional 
regularization scheme \cite{Bastianelli:2000nm}
which requires finite covariant counterterms
only\footnote{Mode regularization 
\cite{Bastianelli:1998jm,Bastianelli:1992be} and time slicing 
\cite{deBoer:1995hv} need instead noncovariant 
counterterms which render higher loop calculations even more laborious.
See ref. \cite{Hatzinikitas:2001pk} for an attempt to computerize
the time slicing procedure and refs. \cite{new}
for the original use of dimensional regularization 
in the infinite propagation time limit.}.
Such a lengthy calculation was indeed performed recently in 
\cite{last}, 
confirming the correctness of the dimensional regularization scheme of the
quantum mechanical path integral \cite{Bastianelli:2000nm}
as well as the correct value of the trace anomalies identified in 
\cite{Bastianelli:2000hi}.

In this letter we wish to discuss a simplified approach to obtain trace 
anomalies in $d\leq6$ (hopefully it  may be extended to higher 
dimensions as well in the future).
The strategy we propose is to take advantage of some recent results
concerning trace anomalies. These results guarantee that one may obtain 
all but one terms in the anomaly by a simpler $n$-loop
calculation in the 1d nonlinear sigma model with arbitrary 
geometry.
Then, the missing part of the anomaly can be identified by
a $(n+1)$-loop calculation performed
in the simpler geometry of a maximally symmetric space. 
It is this geometry which renders the higher loop 
calculation much easier.

Concretely, we first make use of the classification of Deser
and Schwimmer \cite{deser}
that divides trace anomalies into: {\it type A}, which are unique and
always proportional to the Euler topological density,
{\it type B}, whose number increases with the spacetime dimensions
 and are made up by local Weyl invariants, 
and {\it trivial anomalies}, which can be canceled by the variation 
of local counterterms and can be expressed as total derivatives.
This classification makes it evident that 
the simplified geometry of a maximally symmetric space
annihilates type B and trivial anomalies,
and allows a simpler calculation of the type A
anomaly, as done indeed in \cite{Copeland:1986ua}.
In  the path integral method the type A anomaly can be obtained by a 
$(n+1)$-loop calculation for the sigma model on the maximally symmetric 
geometry. The latter simplifies 
drastically the calculation\footnote{It is conceivable that one may 
devise a way of computing this path integral exactly, thus deriving 
a compact formula for a generating functional for type A anomalies.}.
Then, inspecting the cohomological analysis for trace anomalies 
in $d=4$ \cite{Reina} and $d=6$ \cite{Bastianelli:2000rs}
one notices that the remaining non-trivial part
of the trace anomalies (i.e. type B) can be identified 
by certain terms in the curvature that are not affected by adding
trivial anomalies, and at the same time are given 
by disconnected diagrams in the path integral approach. 
The latter are identified by a lower loop calculation (i.e. at $n$-loops). 

Let us consider the case of a conformal scalar field in $d$ dimensions
\bea
I = \int d^d x {\sqrt g}\ {1\over 2} ( g^{\mu\nu}
\partial_\mu \phi \partial_\nu \phi - \xi R \phi^2)
\eea
where $\xi= { d-2 \over 4(d-1)}$
and $R$ is the curvature scalar. Our conventions for the curvature
tensors follow from
$ [\nabla_\mu, \nabla_\nu] V^\rho = R_{\mu\nu}{}^\rho{}_\sigma V^\sigma$
and $R_{\mu\nu} = R_{\mu\sigma}{}^\sigma{}_\nu$.
We employ an euclidean signature.

As described in \cite{Bastianelli:1992be}
one-loop trace anomalies can be obtained by computing a certain 
Fujikawa jacobian suitably regulated and represented as a quantum 
mechanical path integral with periodic boundary conditions
\bea
\int d^d x\ \sqrt{g}\ \sigma(x) \langle T^\mu{}_\mu (x) \rangle=
\lim_{\beta \rightarrow 0}
{\rm Tr} [ \sigma\ {\rm e}^{-\beta H}] = \lim_{\beta \rightarrow 0}
\int_{_{PBC}} \hskip -.6cm {\cal D}x\ \sigma (x)\ {\rm e}^{-S[x]} 
\label{due}
\eea
where on the left hand side $T^\mu{}_\mu$ denotes the trace of the 
stress tensor 
$ T_{\mu\nu} = {2\over \sqrt{g}}{\delta I\over \delta g^{\mu\nu}}$
of the  conformal scalar 
and $\sigma(x)$
is an arbitrary function describing an infinitesimal Weyl variation.
In the first equality the infinitesimal part of the Fujikawa 
jacobian has been regulated with the conformal scalar field
kinetic operator $H = -{1\over 2}\nabla^2 -{\xi \over 2 }R$.
The limit $\beta \rightarrow 0$ should be taken after removing 
divergent terms in $\beta$ (which is what the renormalization in
QFT does). Thus, it picks up just the $\beta$ independent term.
Finally, on the right hand side the trace is given a representation as
a quantum mechanical path integral corresponding to a model with 
hamiltonian $H$ and with periodic boundary conditions (PBC).
Using the dimensional regularization scheme \cite{Bastianelli:2000nm}
the path integral requires the action
\bea
S[x] = {1\over \beta} \int_{-1}^{0} \!\!\! dt\ 
\biggl [ {1\over 2} g_{\mu\nu}(x)\dot x^\mu \dot x^\nu + \beta^2 
[V(x) +V_{DR}(x)] \biggr ]
\label{tre}
\eea
with a scalar potential $V= -{\xi\over 2 }R$ and 
the counterterm $V_{DR}= {1\over 8 }R$, both needed
to reproduce the correct hamiltonian  $H$.
As in \cite{Bastianelli:2000nm}
a ghost action will be used to exponentiate the nontrivial 
part of the path integral measure
\bea
S_{gh} = {1\over \beta} \int_{-1}^{0} \!\!\! dt\ 
\biggl [ {1\over 2} g_{\mu\nu}(x) ( a^\mu  a^\nu +b^\mu  c^\nu )
\biggr ]
\label{gho}
\eea
where $a^\mu$ are bosonic and $b^\mu$, $c^\mu$ fermionic ghosts.

One can easily eliminate the arbitrary function $\sigma(x)$
from (\ref{due}) to obtain the local formula
\bea
\langle T^\mu{}_\mu (x) \rangle=
\lim_{\beta \rightarrow 0}
\int_{_{x(-1)=x(0)=x}} \hskip -1.7cm {\cal D}x\ {\rm e}^{-S[x]} 
\equiv \lim_{\beta \rightarrow 0} Z(\beta)
\eea
where again the limit has to be understood only as the 
indication of extracting the $\beta$ independent part
of the subsequent expression, while the boundary conditions 
on the path integral identify the initial and final points
and keep them fixed.

Aiming at exemplifying the proposed method for 
 $d \leq 6 $ we need the transition amplitude  $Z(\beta)$
on an arbitrary geometry up to three loops. This 
has already been computed and can be read off from various 
papers (see e.g. \cite{Bastianelli:1998jm,last}) 
\bea
Z(\beta) = 
{1\over (2\pi \beta)^{d\over 2}}  
\exp \biggl [ {\beta \over 12} (6 \xi  - 1) R
+\beta^2 \biggl({1\over 720} (R^2_{mnab}-R_{mn}^2)+
{1\over 120} (5 \xi - 1 ) \nabla^2 R  \biggr)
+ O(\beta^3)
\biggr].
\label{formula-general}
\eea

Now we need to compute at 4-loops on
the simplified geometry of a maximally symmetric (MS) space, 
where the Riemann tensor is expressed by
\bea 
R_{mnab} = b (g_{ma}g_{nb}-g_{mb}g_{na})
\eea
with
\bea 
b = {R\over d(1-d)}.
\ena
We find it easier to use Riemann normal coordinates,
so that cubic vertices are absent.
The expansion of the metric in Riemann normal coordinates
around a point (to be called the origin) is easily obtained by the method
explained in \cite{Bastianelli:1992be} and reads
\bea
 g_{mn}(x) dx^m dx^n = 
\biggl[
\delta_{mn} + 2 (x_mx_n - \delta_{mn} \vec{x}^{\, 2})\biggl( {b\over 6}  -
{16\over 6!} b^2 \vec{x}^{\, 2} + {8\over 7!} b^3 (\vec{x}^{\, 2})^2
 +\cdots\biggr)
\biggl] dx^mdx^n.
\label{nove}
\eea
As an aside, we note that it is easy to evaluate recursively
all the terms in the expansion and sum them up in a compact form 
\bea 
g_{mn}(x) dx^m dx^n &=& \vec{dx}^2 + 
2 \sum_{n=1}^\infty {(-4b)^n (\vec{x}^{\, 2})^{n-1}\over (2n+2)!}
(\vec{x}^{\, 2} \vec{dx}^2 -(\vec{x}\cdot\vec{dx})^2 )
\\
&=&\vec{dx}^2 +
{1-2b \vec{x}^{\, 2} -\cos(2 \sqrt{b \vec{x}^{\, 2}})\over 2 b 
(\vec{x}^{\, 2})^2 }
(\vec{x}^{\, 2} \vec{dx}^2 -(\vec{x}\cdot\vec{dx})^2 ).
\eea
Using (\ref{nove}) into (\ref{tre}) and (\ref{gho}) produces the required 
sigma model action. As usual, after getting the free propagators 
from the quadratic part of the action,  one is left to compute
perturbatively using Wick contractions
\bea
\hat 
Z(\beta) = 
{1\over (2\pi \beta)^{d\over 2}}  
\exp \biggl [- \beta (1-4\xi) {R\over 8}
\biggr] \langle {\rm e}^{- \hat S_{int}} \rangle 
\eea
with
\bea 
\hat S_{int} &=& {1\over \beta} \int_{-1}^{0} \!\!\! dt\ 
\biggl(
{b\over 6} - {16\over 6!} b^2 \vec{x}^{\, 2}
+ {8\over 7!} b^3 (\vec{x}^{\, 2})^2
+\cdots \biggr)
(x_mx_n -\delta_{mn}\vec{x}^{\, 2})( \dot x^m \dot x^n + a^m a^n +b^m c^n)
\nonumber
\\
&=& S_4+S_6 +S_8+\cdots
\eea
where the subscripts indicate the power of the quantum 
fields appearing in the given vertex.
Up to the order $\beta^3$ we only need to compute
the following connected diagrams
\bea
\langle {\rm e}^{- S_{int}} \rangle =
\exp \biggl [ 
- \langle S_4 \rangle
- \langle S_6 \rangle
- \langle S_8 \rangle
+ {1\over 2}  \langle S_4^2 \rangle_c
+ \langle S_4 S_6 \rangle_c
- {1\over  6}  \langle S_4^3  \rangle_c
+ O(\beta^{4}) \biggr ].
\label{cd}
\eea
Using Wick contractions and dimensional regularization we find
\EQ
\begin{array}{ll}
 \langle S_4 \rangle = -{\beta\over 4!} R  
& \langle S_6 \rangle =  - {\beta^2\over 5!} 
 {(d+2)\over 9 d (d-1)}R^2 \\ [3mm]
 \langle S_8 \rangle =  - {\beta^3\over 7!} 
 {(d+2)(d+4)\over 15 d^2 (d-1)^2} R^3 &
\langle S_4^2 \rangle_c = - {\beta^2\over 4!} {1\over 9 d} R^2  \\ [3mm]
\langle S_4 S_6 \rangle_c = - {\beta^3\over 6!} 
{4(d+2)\over 45 d^2(d-1)} R^3 \ \ \ \ \ \ \ \ \ \ \ \
& \langle S_4^3 \rangle_c = - {\beta^3\over 6!} 
{2 (d^2-4)\over 3 d^2(d-1)^2} R^3
\label{inv}
\end{array}
\EN 
so that the full answer reads
\bea
\hat 
Z(\beta) = 
{1\over (2\pi \beta)^{d\over 2}}  
\exp \biggl [{\beta\over 4!}(12\xi-2) R
- {\beta^2\over 6!} {(d-3)\over d (d-1)} R^2
+ {\beta^3\over 8!} {16 (d+2)(d-3)\over 9d^2 (d-1)^2} R^3
+ \cdots
\biggr] .
\label{formula-mss}
\eea

We are now ready to describe concretely our method.
In $d=2$ there is only the A type of anomaly. It can 
most easily be computed at two loops (i.e. order $\beta$ in eq. 
(\ref{cd})) in the simplified MS geometry.
Thus, using the two loop part of (\ref{formula-mss}) and setting $d=2$ 
and $\xi=0$ gives
\bea
\langle T^\mu{}_\mu (x) \rangle_2=- {1\over 24 \pi} R
\eea
which is the correct anomaly for a scalar.

In $d=4$ we have instead $\xi={1\over 6}$, and the 3-loop calculation
on the MS geometry 
(i.e. using terms up to order $\beta^2$ in the exponent of 
eq. (\ref{formula-mss})) 
gives
\bea
\langle T^\mu{}_\mu (x) \rangle_4\biggl|_{\rm type A}= - {1\over (2\pi)^2}
{1\over 6!} {1\over 12} R^2 =  - {1\over (2\pi)^2} 
{1\over 6!} {1\over 2} E_4
\eea
where in the second equality we have used the topological Euler density
evaluated on the MS geometry
\bea 
E_4 \equiv R_{mnab}^2 -4 R_{mn}^2 +R^2 = {1\over 6} R^2.
\eea
Now, the most  general expression for $4d$ trace anomalies 
was obtained through a cohomological analysis in \cite{Reina} 
and reads
\bea
\langle T^\mu{}_\mu (x) \rangle_4= {1\over (2\pi)^2} {1\over 6!}
(a\, E_4 + c\, C + d\, \square R)
\label{trace4}
\eea
where  $C$ is the square of the Weyl tensor  
representing the type B anomaly
\bea
C \equiv R_{mnab}^2 -2 R_{mn}^2 +{1\over 3}R^2
\eea
and $ \square R$ is the unique trivial anomaly.
Clearly a modification of the trivial anomaly cannot change 
the coefficient of the $R^2$ term
appearing implicitly in eq. (\ref{trace4}).
At the same time this coefficient can be 
produced by disconnected diagrams only. 
In fact, no index contraction signalling a connection through 
a propagator appears between the two
factors of the curvature tensors belonging each to a separate vertex.
Thus, the coefficient of the $R^2$ term is directly identified by a 
lower loop calculation, namely a
2-loop calculation on an arbitrary geometry (order $\beta$). 
Said differently, it is clear from eq. (\ref{formula-general}) 
that for generic manifolds
an explicit $R^2$ dependence can only come from
the linear $R$ terms in the exponent. In $d=4$ this term is
absent. Since this can not come from trivial
anomalies one can conclude that
\bea
c= 3(r-a)
\label{dim4}
\eea
and hence $c={3\over2}$. Thus, the complete trace anomaly for a 
$ d=4$ conformal scalar reads
\bea
\langle T^\mu{}_\mu (x) \rangle_4= {1\over (2\pi)^2} {1\over 6!} 
\biggl(- {1\over 2} E_4 +  {3\over 2} C \biggr).
\label{finaltrace}
\eea

Now, let's address the $d=6$ case. In this spacetime dimensions
$\xi={1\over 5}$
and the 4-loop computation (order $\beta^3$)
on the MS geometry presented above produces
\bea
\langle T^\mu{}_\mu (x) \rangle_6\biggl|_{\rm type A}=
- {1\over (2\pi)^3} {1\over 8!}
{2\over 135} R^3 =  - {1\over (2\pi)^3} {1\over 8!}{5\over 72} E_6
\eea
where in the last equality we have used the following topological density
evaluated on the MS geometry
\bea
E_{6} \equiv -\epsilon_{m_1n_1m_2n_2m_3n_3}\epsilon^{a_1b_1a_2b_2a_3b_3}
R^{m_1n_1}{}_{a_1b_1}R^{m_2n_2}{}_{a_2b_2} R^{m_3n_3}{}_{a_3b_3}
= {16\over 75}R^3 .
\ena
The general expression of 
6d trace anomalies can be obtained from a cohomological analysis
\cite{bon, Bastianelli:2000rs}
and is of the form 
\bea
\langle T^a{}_a \rangle = {1\over (2\pi)^3} {1\over 8!}
(a\, E_6 +c_1\, I_1 +c_2\, I_2 + c_3\, I_3 +
{\rm trivial \ anomalies})
\eea
with the three Weyl invariants given by
\bea
I_{1} &=& C_{amnb} C^{mijn} C_{i}{}^{ab}{}_j \nonumber \\
I_{2} &=& C_{ab}{}^{mn} C_{mn}{}^{ij} C_{ij}{}^{ab} \\
I_{3} &=& C_{mabc}\left(\nabla^{2}\delta^{m}_n-4R^{m}_n
+\frac{6}{5}R\delta^{m}_n\right)C^{nabc}+{\rm trivial \ anomalies}. 
\nonumber
\ena
where $ C_{abmn}$ is the Weyl tensor in 6 dimensions and
the coefficients $a,c_1,c_2,c_3$ will depend on the particular
model considered.
Now, it is important to recall that the  
specific expressions of the trivial anomalies have been
recently found in \cite{Bastianelli:2000rs}.
Consulting those results (see, in particular, table I) 
one notice that the 
coefficients of the three structures $ R^3, R R_{mn}^2, RR_{mnab}^2$
can never be modified by adding trivial anomalies.
At the same time those structures
can be obtained by disconnected diagrams only.
Thus, it suffices to use the results at the 3rd-loop order
on an arbitrary geometry to fix them. 

For the case of the conformal scalar field we use 
eq. (\ref{formula-general}) and obtain
\bea
\langle T^\mu{}_\mu (x) \rangle_6 = {1\over (2\pi)^3} {1\over 8!}
(r_1\, R^3 + r_2\, R R_{mn}^2+r_3\, RR_{mnab}^2
+ {\rm other\ structures})
\eea
with $r_1={7\over 225}$ and $r_3=-r_2={14\over 15}$.
Since these coefficients must correspond to the sum of the 
type A and B anomalies only, one finds with  simple linear algebra
\bea
\langle T^a{}_a \rangle &=& {1\over (2\pi)^3} {1\over 8!}
\biggl( -{5\over 72}E_6 -{28\over 3} I_1 +{5\over 3}I_2 + 2I_3
\biggr).
\eea
The general formulas relating the various coefficients are
\bea
c_1 &=& {4\over 3} (168 a +4 r_2+9 r_3) \nonumber \\
c_2 &=& {1\over 3} (-408 a + 300 r_1+ 16 r_2 - 19 r_3) 
\label{dim6}
\\
c_3 &=& -{5\over 3} (24 a - 15 r_1 - r_2) . \nonumber
\eea

Before closing, it is amusing to note that on the three sphere
$S^3$ (i.e. on the group manifold of $SU(2)$)
the higher order corrections in (\ref{formula-mss}) vanish.
In fact, the transition amplitude on $S^3$ is 
known exactly \cite{Schulman}. From our path integral perspective
this amplitude is saturated by the two loop correction given in eq. 
(\ref{formula-mss}).

We have discussed a simplified method for computing
trace anomalies in  $d \leq 6 $. Apart from some amusing 
path integral computations on maximally symmetric spaces,
our main results are (\ref{dim4}) and (\ref{dim6}).
It would be interesting to extend this method to higher
dimension. The main problem is to analyze how
trivial anomalies  may affect the structure of 
type B anomalies in $d \geq 8 $. 
This is at present unknown.
On the other hand, it is fortunate that most recent applications,
such as the search for a $C$-theorem in higher dimensions, 
concern mostly the type A anomalies.

\vfill\eject

\end{document}